\documentclass[conference,10pt,twoside,twocolumn]{IEEEtran}
\IEEEoverridecommandlockouts

\usepackage{cite}
\usepackage[T1]{fontenc}
\usepackage{graphicx}
\usepackage{amssymb}
\usepackage{amsmath}
\usepackage{amsthm}
\usepackage{subfigure}
\usepackage{booktabs} 
\usepackage{multirow}
\usepackage{microtype}
\usepackage{balance}
\usepackage{xcolor}

\usepackage[hidelinks]{hyperref}

\usepackage{amssymb}
\usepackage{amsfonts}
\usepackage{mathrsfs}
\usepackage{xspace}
\usepackage{bm}
\usepackage{upgreek}

\newcommand{\safemath}[2]{\newcommand{#1}{\ensuremath{#2}\xspace}}

\safemath{\bma}{\mathbf{a}}
\safemath{\bmb}{\mathbf{b}}
\safemath{\bmc}{\mathbf{c}}
\safemath{\bmd}{\mathbf{d}}
\safemath{\bme}{\mathbf{e}}
\safemath{\bmf}{\mathbf{f}}
\safemath{\bmg}{\mathbf{g}}
\safemath{\bmh}{\mathbf{h}}
\safemath{\bmi}{\mathbf{i}}
\safemath{\bmj}{\mathbf{j}}
\safemath{\bmk}{\mathbf{k}}
\safemath{\bml}{\mathbf{l}}
\safemath{\bmm}{\mathbf{m}}
\safemath{\bmn}{\mathbf{n}}
\safemath{\bmo}{\mathbf{o}}
\safemath{\bmp}{\mathbf{p}}
\safemath{\bmq}{\mathbf{q}}
\safemath{\bmr}{\mathbf{r}}
\safemath{\bms}{\mathbf{s}}
\safemath{\bmt}{\mathbf{t}}
\safemath{\bmu}{\mathbf{u}}
\safemath{\bmv}{\mathbf{v}}
\safemath{\bmw}{\mathbf{w}}
\safemath{\bmx}{\mathbf{x}}
\safemath{\bmy}{\mathbf{y}}
\safemath{\bmz}{\mathbf{z}}
\safemath{\bmzero}{\mathbf{0}}
\safemath{\bmone}{\mathbf{1}}

\bmdefine{\biad}{a}
\bmdefine{\bibd}{b}
\bmdefine{\bicd}{c}
\bmdefine{\bidd}{d}
\bmdefine{\bied}{e}
\bmdefine{\bifd}{f}
\bmdefine{\bigd}{g}
\bmdefine{\bihd}{h}
\bmdefine{\biid}{i}
\bmdefine{\bijd}{j}
\bmdefine{\bikd}{k}
\bmdefine{\bild}{l}
\bmdefine{\bimd}{m}
\bmdefine{\bind}{n}
\bmdefine{\biod}{o}
\bmdefine{\bipd}{p}
\bmdefine{\biqd}{q}
\bmdefine{\bird}{r}
\bmdefine{\bisd}{s}
\bmdefine{\bitd}{t}
\bmdefine{\biud}{u}
\bmdefine{\bivd}{v}
\bmdefine{\biwd}{w}
\bmdefine{\bixd}{x}
\bmdefine{\biyd}{y}
\bmdefine{\bizd}{z}

\bmdefine{\bixid}{\xi}
\bmdefine{\bilambdad}{\lambda}
\bmdefine{\bimud}{\mu}
\bmdefine{\bithetad}{\theta}
\bmdefine{\biphid}{\phi}
\bmdefine{\bideltad}{\delta}

\safemath{\bmia}{\biad}
\safemath{\bmib}{\bibd}
\safemath{\bmic}{\bicd}
\safemath{\bmid}{\bidd}
\safemath{\bmie}{\bied}
\safemath{\bmif}{\bifd}
\safemath{\bmig}{\bigd}
\safemath{\bmih}{\bihd}
\safemath{\bmii}{\biid}
\safemath{\bmij}{\bijd}
\safemath{\bmik}{\bikd}
\safemath{\bmil}{\bild}
\safemath{\bmim}{\bimd}
\safemath{\bmin}{\bind}
\safemath{\bmio}{\biod}
\safemath{\bmip}{\bipd}
\safemath{\bmiq}{\biqd}
\safemath{\bmir}{\bird}
\safemath{\bmis}{\bisd}
\safemath{\bmit}{\bitd}
\safemath{\bmiu}{\biud}
\safemath{\bmiv}{\bivd}
\safemath{\bmiw}{\biwd}
\safemath{\bmix}{\bixd}
\safemath{\bmiy}{\biyd}
\safemath{\bmiz}{\bizd}

\safemath{\bmxi}{\bixid}
\safemath{\bmlambda}{\bilambdad}
\safemath{\bmmu}{\bimud}
\safemath{\bmtheta}{\bithetad}
\safemath{\bmphi}{\biphid}
\safemath{\bmdelta}{\bideltad}

\safemath{\bA}{\mathbf{A}}
\safemath{\bB}{\mathbf{B}}
\safemath{\bC}{\mathbf{C}}
\safemath{\bD}{\mathbf{D}}
\safemath{\bE}{\mathbf{E}}
\safemath{\bF}{\mathbf{F}}
\safemath{\bG}{\mathbf{G}}
\safemath{\bH}{\mathbf{H}}
\safemath{\bI}{\mathbf{I}}
\safemath{\bJ}{\mathbf{J}}
\safemath{\bK}{\mathbf{K}}
\safemath{\bL}{\mathbf{L}}
\safemath{\bM}{\mathbf{M}}
\safemath{\bN}{\mathbf{N}}
\safemath{\bO}{\mathbf{O}}
\safemath{\bP}{\mathbf{P}}
\safemath{\bQ}{\mathbf{Q}}
\safemath{\bR}{\mathbf{R}}
\safemath{\bS}{\mathbf{S}}
\safemath{\bT}{\mathbf{T}}
\safemath{\bU}{\mathbf{U}}
\safemath{\bV}{\mathbf{V}}
\safemath{\bW}{\mathbf{W}}
\safemath{\bX}{\mathbf{X}}
\safemath{\bY}{\mathbf{Y}}
\safemath{\bZ}{\mathbf{Z}}

\safemath{\bZero}{\mathbf{0}}
\safemath{\bOne}{\mathbf{1}}
\safemath{\bDelta}{\mathbf{\Delta}}
\safemath{\bLambda}{\mathbf{\Lambda}}
\safemath{\bPhi}{\mathbf{\Upphi}}
\safemath{\bSigma}{\mathbf{\Upsigma}}
\safemath{\bOmega}{\mathbf{\Upomega}}
\safemath{\bTheta}{\mathbf{\Uptheta}}

\bmdefine{\biAd}{A}
\bmdefine{\biBd}{B}
\bmdefine{\biCd}{C}
\bmdefine{\biDd}{D}
\bmdefine{\biEd}{E}
\bmdefine{\biFd}{F}
\bmdefine{\biGd}{G}
\bmdefine{\biHd}{H}
\bmdefine{\biId}{I}
\bmdefine{\biJd}{J}
\bmdefine{\biKd}{K}
\bmdefine{\biLd}{L}
\bmdefine{\biMd}{M}
\bmdefine{\biOd}{N}
\bmdefine{\biPd}{O}
\bmdefine{\biQd}{P}
\bmdefine{\biRd}{R}
\bmdefine{\biSd}{S}
\bmdefine{\biTd}{T}
\bmdefine{\biUd}{U}
\bmdefine{\biVd}{V}
\bmdefine{\biWd}{W}
\bmdefine{\biXd}{X}
\bmdefine{\biYd}{Y}
\bmdefine{\biZd}{Z}

\bmdefine{\biDelta}{\Delta}
\bmdefine{\biLambda}{\Lambda}
\bmdefine{\biPhi}{\Phi}
\bmdefine{\biSigma}{\Sigma}
\bmdefine{\biOmega}{\Omega}
\bmdefine{\biTheta}{\Theta}

\safemath{\bimA}{\biAd}
\safemath{\bimB}{\biBd}
\safemath{\bimC}{\biCd}
\safemath{\bimD}{\biDd}
\safemath{\bimE}{\biEd}
\safemath{\bimF}{\biFd}
\safemath{\bimG}{\biGd}
\safemath{\bimH}{\biHd}
\safemath{\bimI}{\biId}
\safemath{\bimJ}{\biJd}
\safemath{\bimK}{\biKd}
\safemath{\bimL}{\biLd}
\safemath{\bimM}{\biMd}
\safemath{\bimN}{\biNd}
\safemath{\bimO}{\biOd}
\safemath{\bimP}{\biPd}
\safemath{\bimQ}{\biQd}
\safemath{\bimR}{\biRd}
\safemath{\bimS}{\biSd}
\safemath{\bimT}{\biTd}
\safemath{\bimU}{\biUd}
\safemath{\bimV}{\biVd}
\safemath{\bimW}{\biWd}
\safemath{\bimX}{\biXd}
\safemath{\bimY}{\biYd}
\safemath{\bimZ}{\biZd}

\safemath{\bimDelta}{\biDelta}
\safemath{\bimLambda}{\biLambda}
\safemath{\bimPhi}{\biPhi}
\safemath{\bimSigma}{\biSigma}
\safemath{\bimOmega}{\biOmega}
\safemath{\bimTheta}{\biTheta}

\safemath{\setA}{\mathcal{A}}
\safemath{\setB}{\mathcal{B}}
\safemath{\setC}{\mathcal{C}}
\safemath{\setD}{\mathcal{D}}
\safemath{\setE}{\mathcal{E}}
\safemath{\setF}{\mathcal{F}}
\safemath{\setG}{\mathcal{G}}
\safemath{\setH}{\mathcal{H}}
\safemath{\setI}{\mathcal{I}}
\safemath{\setJ}{\mathcal{J}}
\safemath{\setK}{\mathcal{K}}
\safemath{\setL}{\mathcal{L}}
\safemath{\setM}{\mathcal{M}}
\safemath{\setN}{\mathcal{N}}
\safemath{\setO}{\mathcal{O}}
\safemath{\setP}{\mathcal{P}}
\safemath{\setQ}{\mathcal{Q}}
\safemath{\setR}{\mathcal{R}}
\safemath{\setS}{\mathcal{S}}
\safemath{\setT}{\mathcal{T}}
\safemath{\setU}{\mathcal{U}}
\safemath{\setV}{\mathcal{V}}
\safemath{\setW}{\mathcal{W}}
\safemath{\setX}{\mathcal{X}}
\safemath{\setY}{\mathcal{Y}}
\safemath{\setZ}{\mathcal{Z}}
\safemath{\emptySet}{\varnothing}

\safemath{\colA}{\mathscr{A}}
\safemath{\colB}{\mathscr{B}}
\safemath{\colC}{\mathscr{C}}
\safemath{\colD}{\mathscr{D}}
\safemath{\colE}{\mathscr{E}}
\safemath{\colF}{\mathscr{F}}
\safemath{\colG}{\mathscr{G}}
\safemath{\colH}{\mathscr{H}}
\safemath{\colI}{\mathscr{I}}
\safemath{\colJ}{\mathscr{J}}
\safemath{\colK}{\mathscr{K}}
\safemath{\colL}{\mathscr{L}}
\safemath{\colM}{\mathscr{M}}
\safemath{\colN}{\mathscr{N}}
\safemath{\colO}{\mathscr{O}}
\safemath{\colP}{\mathscr{P}}
\safemath{\colQ}{\mathscr{Q}}
\safemath{\colR}{\mathscr{R}}
\safemath{\colS}{\mathscr{S}}
\safemath{\colT}{\mathscr{T}}
\safemath{\colU}{\mathscr{U}}
\safemath{\colV}{\mathscr{V}}
\safemath{\colW}{\mathscr{W}}
\safemath{\colX}{\mathscr{X}}
\safemath{\colY}{\mathscr{Y}}
\safemath{\colZ}{\mathscr{Z}}

\safemath{\opA}{\mathbb{A}}
\safemath{\opB}{\mathbb{B}}
\safemath{\opC}{\mathbb{C}}
\safemath{\opD}{\mathbb{D}}
\safemath{\opE}{\mathbb{E}}
\safemath{\opF}{\mathbb{F}}
\safemath{\opG}{\mathbb{G}}
\safemath{\opH}{\mathbb{H}}
\safemath{\opI}{\mathbb{I}}
\safemath{\opJ}{\mathbb{J}}
\safemath{\opK}{\mathbb{K}}
\safemath{\opL}{\mathbb{L}}
\safemath{\opM}{\mathbb{M}}
\safemath{\opN}{\mathbb{N}}
\safemath{\opO}{\mathbb{O}}
\safemath{\opP}{\mathbb{P}}
\safemath{\opQ}{\mathbb{Q}}
\safemath{\opR}{\mathbb{R}}
\safemath{\opS}{\mathbb{S}}
\safemath{\opT}{\mathbb{T}}
\safemath{\opU}{\mathbb{U}}
\safemath{\opV}{\mathbb{V}}
\safemath{\opW}{\mathbb{W}}
\safemath{\opX}{\mathbb{X}}
\safemath{\opY}{\mathbb{Y}}
\safemath{\opZ}{\mathbb{Z}}
\safemath{\opZero}{\mathbb{O}}
\safemath{\identityop}{\opI}

\safemath{\veca}{\bma}
\safemath{\vecb}{\bmb}
\safemath{\vecc}{\bmc}
\safemath{\vecd}{\bmd}
\safemath{\vece}{\bme}
\safemath{\vecf}{\bmf}
\safemath{\vecg}{\bmg}
\safemath{\vech}{\bmh}
\safemath{\veci}{\bmi}
\safemath{\vecj}{\bmj}
\safemath{\veck}{\bmk}
\safemath{\vecl}{\bml}
\safemath{\vecm}{\bmm}
\safemath{\vecn}{\bmn}
\safemath{\veco}{\bmo}
\safemath{\vecp}{\bmp}
\safemath{\vecq}{\bmq}
\safemath{\vecr}{\bmr}
\safemath{\vecs}{\bms}
\safemath{\vect}{\bmt}
\safemath{\vecu}{\bmu}
\safemath{\vecv}{\bmv}
\safemath{\vecw}{\bmw}
\safemath{\vecx}{\bmx}
\safemath{\vecy}{\bmy}
\safemath{\vecz}{\bmz}

\safemath{\veczero}{\bmzero}
\safemath{\vecone}{\bmone}
\safemath{\vecxi}{\bmxi}
\safemath{\veclambda}{\bmlambda}
\safemath{\vecmu}{\bmmu}
\safemath{\vectheta}{\bmtheta}
\safemath{\vecphi}{\bmphi}
\safemath{\vecdelta}{\bmdelta}

\safemath{\matA}{\bA}
\safemath{\matB}{\bB}
\safemath{\matC}{\bC}
\safemath{\matD}{\bD}
\safemath{\matE}{\bE}
\safemath{\matF}{\bF}
\safemath{\matG}{\bG}
\safemath{\matH}{\bH}
\safemath{\matI}{\bI}
\safemath{\matJ}{\bJ}
\safemath{\matK}{\bK}
\safemath{\matL}{\bL}
\safemath{\matM}{\bM}
\safemath{\matN}{\bN}
\safemath{\matO}{\bO}
\safemath{\matP}{\bP}
\safemath{\matQ}{\bQ}
\safemath{\matR}{\bR}
\safemath{\matS}{\bS}
\safemath{\matT}{\bT}
\safemath{\matU}{\bU}
\safemath{\matV}{\bV}
\safemath{\matW}{\bW}
\safemath{\matX}{\bX}
\safemath{\matY}{\bY}
\safemath{\matZ}{\bZ}
\safemath{\matzero}{\bmzero}

\safemath{\matDelta}{\bDelta}
\safemath{\matLambda}{\bLambda}
\safemath{\matPhi}{\bPhi}
\safemath{\matSigma}{\bSigma}
\safemath{\matOmega}{\bOmega}
\safemath{\matTheta}{\bTheta}

\safemath{\matidentity}{\matI}
\safemath{\matone}{\matO}

\safemath{\rnda}{A}
\safemath{\rndb}{B}
\safemath{\rndc}{C}
\safemath{\rndd}{D}
\safemath{\rnde}{E}
\safemath{\rndf}{F}
\safemath{\rndg}{G}
\safemath{\rndh}{H}
\safemath{\rndi}{I}
\safemath{\rndj}{J}
\safemath{\rndk}{K}
\safemath{\rndl}{L}
\safemath{\rndm}{M}
\safemath{\rndn}{N}
\safemath{\rndo}{O}
\safemath{\rndp}{P}
\safemath{\rndq}{Q}
\safemath{\rndr}{R}
\safemath{\rnds}{S}
\safemath{\rndt}{T}
\safemath{\rndu}{U}
\safemath{\rndv}{V}
\safemath{\rndw}{W}
\safemath{\rndx}{X}
\safemath{\rndy}{Y}
\safemath{\rndz}{Z}

\safemath{\rveca}{\bimA}
\safemath{\rvecb}{\bimB}
\safemath{\rvecc}{\bimC}
\safemath{\rvecd}{\bimD}
\safemath{\rvece}{\bimE}
\safemath{\rvecf}{\bimF}
\safemath{\rvecg}{\bimG}
\safemath{\rvech}{\bimH}
\safemath{\rveci}{\bimI}
\safemath{\rvecj}{\bimJ}
\safemath{\rveck}{\bimK}
\safemath{\rvecl}{\bimL}
\safemath{\rvecm}{\bimM}
\safemath{\rvecn}{\bimN}
\safemath{\rveco}{\bomO}
\safemath{\rvecp}{\bimP}
\safemath{\rvecq}{\bimQ}
\safemath{\rvecr}{\bimR}
\safemath{\rvecs}{\bimS}
\safemath{\rvect}{\bimT}
\safemath{\rvecu}{\bimU}
\safemath{\rvecv}{\bimV}
\safemath{\rvecw}{\bimW}
\safemath{\rvecx}{\bimX}
\safemath{\rvecy}{\bimY}
\safemath{\rvecz}{\bimZ}

\safemath{\rvecxi}{\bmxi}
\safemath{\rveclambda}{\bmlambda}
\safemath{\rvecmu}{\bmmu}
\safemath{\rvectheta}{\bmtheta}
\safemath{\rvecphi}{\bmphi}

\safemath{\rmatA}{\bimA}
\safemath{\rmatB}{\bimB}
\safemath{\rmatC}{\bimC}
\safemath{\rmatD}{\bimD}
\safemath{\rmatE}{\bimE}
\safemath{\rmatF}{\bimF}
\safemath{\rmatG}{\bimG}
\safemath{\rmatH}{\bimH}
\safemath{\rmatI}{\bimI}
\safemath{\rmatJ}{\bimJ}
\safemath{\rmatK}{\bimK}
\safemath{\rmatL}{\bimL}
\safemath{\rmatM}{\bimM}
\safemath{\rmatN}{\bimN}
\safemath{\rmatO}{\bimO}
\safemath{\rmatP}{\bimP}
\safemath{\rmatQ}{\bimQ}
\safemath{\rmatR}{\bimR}
\safemath{\rmatS}{\bimS}
\safemath{\rmatT}{\bimT}
\safemath{\rmatU}{\bimU}
\safemath{\rmatV}{\bimV}
\safemath{\rmatW}{\bimW}
\safemath{\rmatX}{\bimX}
\safemath{\rmatY}{\bimY}
\safemath{\rmatZ}{\bimZ}

\safemath{\rmatDelta}{\bimDelta}
\safemath{\rmatLambda}{\bimLambda}
\safemath{\rmatPhi}{\bimPhi}
\safemath{\rmatSigma}{\bimSigma}
\safemath{\rmatOmega}{\bimOmega}
\safemath{\rmatTheta}{\bimTheta}

\usepackage{amssymb}
\usepackage{amsfonts}
\usepackage{mathrsfs}
\usepackage{xspace}
\usepackage{bm}
\usepackage{fancyref}
\usepackage{textcomp}

\usepackage{multirow}
\usepackage{stmaryrd}

\newenvironment{textbmatrix}{	\setlength{\arraycolsep}{2.5pt}%
								\big[\begin{matrix}}{\end{matrix}\big]%
								\raisebox{0.08ex}{\vphantom{M}}}

\def\be{\begin{equation}}
\def\ee{\end{equation}}
\def\een{\nonumber \end{equation}}
\def\mat{\begin{bmatrix}}
\def\emat{\end{bmatrix}}
\def\btm{\begin{textbmatrix}}
\def\etm{\end{textbmatrix}}

\def\ba#1\ea{\begin{align}#1\end{align}}
\def\bas#1\eas{\begin{align*}#1\end{align*}}
\def\bs#1\es{\begin{split}#1\end{split}}
\def\bg#1\eg{\begin{gather}#1\end{gather}}
\def\bml#1\eml{\begin{multline}#1\end{multline}}
\def\bi#1\ei{\begin{itemize}#1\end{itemize}}

\newcommand{\lefto}{\mathopen{}\left}

\DeclareMathOperator*{\argmin}{arg\;min}		
\DeclareMathOperator*{\argmax}{arg\;max}		

\newcommand{\vecnorm}[1]{\lefto\lVert#1\right\rVert}		
\newcommand{\frobnorm}[1]{\vecnorm{#1}_{\text{F}}}	
\newcommand{\est}[1]{\ensuremath{\hat{#1}}} 	

\safemath{\dirac}{\delta}					
\safemath{\krond}{\dirac}					

\safemath{\upto}{\uparrow}
\safemath{\downto}{\downarrow}
\safemath{\iu}{j}							
\safemath{\ev}{\lambda}						
\safemath{\hilseqspace}{l^{2}}				
\newcommand{\banachfunspace}[1]{\setL^{#1}}	
\safemath{\hilfunspace}{\banachfunspace{2}}	

\safemath{\SNR}{\textit{SNR}} 				
\safemath{\PAR}{\textit{PAR}} 				
\safemath{\No}{N_0}							
\safemath{\Es}{E_s}							
\safemath{\Eb}{E_b}							
\safemath{\EbNo}{\frac{\Eb}{\No}}
\safemath{\EsNo}{\frac{\Es}{\No}}

\DeclareMathOperator{\CHop}{\ensuremath{\opH}} 
\safemath{\tvir}{\rndh_{\CHop}}				
\safemath{\tvtf}{\rndl_{\CHop}}				
\safemath{\spf}{\rnds_{\CHop}}				
\safemath{\bff}{H_{\CHop}}					

\safemath{\ircf}{r_{h}}						
\safemath{\tftvcf}{r_{s}}					
\safemath{\tfcf}{r_{l}}						
\safemath{\bfcf}{r_{H}}						

\safemath{\tcorr}{c_h}						
\safemath{\scf}{c_{s}}						
\safemath{\tfcorr}{c_{l}}					
\safemath{\fcorr}{c_{H}}						

\safemath{\mi}{I}							
\safemath{\capacity}{C}						

\safemath{\normal}{\mathcal{N}}			
\safemath{\jpg}{\mathcal{CN}}			
\safemath{\mchain}{\leftrightarrow}		

\safemath{\dB}{\,\mathrm{dB}}
\safemath{\dBm}{\,\mathrm{dBm}}
\safemath{\Hz}{\,\mathrm{Hz}}
\safemath{\kHz}{\,\mathrm{kHz}}
\safemath{\MHz}{\,\mathrm{MHz}}
\safemath{\GHz}{\,\mathrm{GHz}}
\safemath{\s}{\,\mathrm{s}}
\safemath{\ms}{\,\mathrm{ms}}
\safemath{\mus}{\,\mathrm{\text{\textmu}s}}
\safemath{\ns}{\,\mathrm{ns}}
\safemath{\ps}{\,\mathrm{ps}}
\safemath{\meter}{\,\mathrm{m}}
\safemath{\mm}{\,\mathrm{mm}}
\safemath{\cm}{\,\mathrm{cm}}
\safemath{\m}{\,\mathrm{m}}
\safemath{\W}{\,\mathrm{W}}
\safemath{\mW}{\, \mathrm{mW}}
\safemath{\J}{\,\mathrm{J}}
\safemath{\K}{\,\mathrm{K}}
\safemath{\bit}{\,\mathrm{bit}}
\safemath{\nat}{\,\mathrm{nat}}

\safemath{\define}{\triangleq}			

\safemath{\equivalent}{\sim}
\safemath{\distas}{\sim}					
\safemath{\sdiff}{\Delta}				

\safemath{\reals}{\mathbb{R}}
\safemath{\positivereals}{\reals_{+}}
\safemath{\integers}{\mathbb{Z}}
\safemath{\posint}{\integers_{+}}
\safemath{\naturals}{\mathbb{N}}
\safemath{\posnaturals}{\naturals_{+}}
\safemath{\complexset}{\mathbb{C}}
\safemath{\rationals}{\mathbb{Q}}

\newcommand*{\fancyrefapplabelprefix}{app}		
\newcommand*{\fancyrefthmlabelprefix}{thm}		
\newcommand*{\fancyreflemlabelprefix}{lem}		
\newcommand*{\fancyrefcorlabelprefix}{cor}		
\newcommand*{\fancyrefdeflabelprefix}{def}		
\newcommand*{\fancyrefproplabelprefix}{prop}		
\newcommand*{\fancyrefexmpllabelprefix}{exmpl}
\newcommand*{\fancyrefalglabelprefix}{alg}		
\newcommand*{\fancyreftbllabelprefix}{tbl}		

\frefformat{vario}{\fancyrefseclabelprefix}{Section~#1}
\frefformat{vario}{\fancyrefthmlabelprefix}{Theorem~#1}
\frefformat{vario}{\fancyreftbllabelprefix}{Table~#1}
\frefformat{vario}{\fancyreflemlabelprefix}{Lemma~#1}
\frefformat{vario}{\fancyrefcorlabelprefix}{Corollary~#1}
\frefformat{vario}{\fancyrefdeflabelprefix}{Definition~#1}
\frefformat{vario}{\fancyreffiglabelprefix}{Fig.~#1}
\frefformat{vario}{\fancyrefapplabelprefix}{Appendix~#1}
\frefformat{vario}{\fancyrefeqlabelprefix}{(#1)}
\frefformat{vario}{\fancyrefproplabelprefix}{Proposition~#1}
\frefformat{vario}{\fancyrefexmpllabelprefix}{Example~#1}
\frefformat{vario}{\fancyrefalglabelprefix}{Algorithm~#1}

\safemath{\nAPm}{n\sb{\textit{AP}}} 			
\safemath{\nUEm}{n\sb{\textit{UE}}} 
\safemath{\nPilot}{n\sb{\textit{pilot}}} 
\safemath{\nCB}{n\sb{\textit{CB}}} 
\safemath{\complexnormal}{\mathcal{CN}}			

\safemath{\dictab}{[\,\dicta\,\,\dictb\,]}

\safemath{\ysig}{\bmy}
\safemath{\ysighat}{\hat{\ysig}}
\safemath{\ysigdim}{M}
\safemath{\xsig}{\bmx}
\safemath{\xsigdim}{N}
\safemath{\nx}{n_x}
\safemath{\zsig}{\bmz}
\safemath{\zsigdim}{\ysigdim}
\safemath{\rsig}{\bmr}
\safemath{\Adict}{\bA}
\safemath{\Adicttilde}{\widetilde{\Adict}}
\safemath{\Adictdim}{\outputdim\times\xsigdim}
\safemath{\avec}{\bma}
\safemath{\avectilde}{\tilde{\avec}}
\safemath{\Bdict}{\bB}
\safemath{\Bdicttilde}{\widetilde{\Bdict}}
\safemath{\Cdict}{\bC}
\safemath{\cvec}{\bmc}
\safemath{\Ddict}{\bD}
\safemath{\Ddictdim}{\ysigdim\times\xsigdim}
\safemath{\dvec}{\bmd}
\safemath{\Ddicttilde}{\widetilde{\bD}}
\safemath{\Bonb}{\bB}
\safemath{\bvec}{\bmb}
\safemath{\Bonbdim}{\ysigdim\times\ysigdim}
\safemath{\noise}{\bmn}
\safemath{\noisedim}{\ysigim}
\safemath{\err}{\bme}
\safemath{\errdim}{\ysigdim}
\safemath{\errset}{\setE}
\safemath{\nerr}{n_e}
\safemath{\delop}{\bP_\errset}
\safemath{\delopc}{\bP_{{\errset}^c}}

\safemath{\cplxi}{\imath}
\safemath{\cplxj}{\jmath}

\safemath{\dict}{\matD}
\safemath{\inputdim}{N}		
\safemath{\outputdim}{M}		
\safemath{\sparsity}{S}	
\safemath{\inputdimA}{{N_a}}	
\safemath{\inputdimB}{{N_b}}	
\safemath{\elemA}{{n_a}}	
\safemath{\elemB}{{n_b}}	
\safemath{\resA}{\matR_a}	
\safemath{\resB}{\matR_b}	
\safemath{\subD}{\matS} 
\safemath{\subA}{\matS_a} 
\safemath{\subB}{\matS_b} 
\safemath{\dicta}{\matA} 	
\safemath{\dictb}{\matB} 	
\safemath{\hollowS}{H}
\safemath{\hollowA}{H_a}
\safemath{\hollowB}{H_b}
\safemath{\cross}{Z}
\safemath{\coh}{\mu_d}			
\safemath{\coha}{\mu_a}			
\safemath{\cohb}{\mu_b}			
\safemath{\mubs}{\nu}	
\safemath{\cohm}{\mu_m} 
\safemath{\dictset}{\setD}	
\safemath{\dictsetp}{\dictset(\coh,\coha,\cohb)}	
\safemath{\dictsetgen}{\dictset_\text{gen}}
\safemath{\dictsetgenp}{\dictsetgen(\coh)}
\safemath{\dictsetonb}{\dictset_\text{onb}}
\safemath{\dictsetonbp}{\dictsetonb(\coh)}

\safemath{\leftside}{U}
\safemath{\rightsideA}{R_a}
\safemath{\rightsideB}{R_b}

\safemath{\indexS}{\setI_S} 

\safemath{\na}{n_a}			
\safemath{\nb}{n_b}			
\safemath{\coeffa}{p_i}	
\safemath{\coeffb}{q_j}	
\safemath{\seta}{\setP}		
\safemath{\setb}{\setQ}     
\safemath{\setw}{\setW}	
\safemath{\setz}{\setZ}	
\safemath{\cola}{\veca}		
\safemath{\colb}{\vecb}		
\safemath{\cold}{\vecd}		
\safemath{\inputvec}{\vecx} 	
\safemath{\error}{\vece}	
\safemath{\noiseout}{\vecz} 	
\safemath{\inputvecel}{x}
\safemath{\inputveca}{\vecx_a}
\safemath{\inputvecb}{\vecx_b}
\safemath{\outputvec}{\vecy}	
\safemath{\lambdamin}{\lambda_{\mathrm{min}}}

\safemath{\elltwo}{\ell_2}
\safemath{\ellone}{\ell_1}
\safemath{\ellzero}{\ell_0}
\safemath{\ellinf}{\ell_\infty}
\safemath{\ellinftilde}{\ell_{\widetilde\infty}}
\safemath{\licard}{Z(\coh,\coha,\cohb)}
\safemath{\xsol}{\hat{x}}
\safemath{\xbord}{x_b}		
\safemath{\xstat}{x_s}		
\safemath{\xstatLone}{\tilde{x}_s}
\safemath{\order}{\mathcal{O}} 
\safemath{\scales}{\Theta} 
\safemath{\ones}{\mathbf{1}} 
\safemath{\zeroes}{\mathbf{0}} 
\safemath{\thlone}{\kappa(\coh,\cohb)} 
\safemath{\constoneA}{\delta} 
\safemath{\constoneB}{\epsilon} 
\safemath{\nlarge}{L}				   
\safemath{\sumlarge}{S_\nlarge}
\safemath{\maxlarger}{P_\nlarge}	   
\safemath{\Pzero}{\textrm{P0}}	
\safemath{\Pone}{\textrm{P1}}
\safemath{\vecfir}{\vecw}			 
\safemath{\vecsec}{\vecz}
\safemath{\elvecfir}{w}              
\safemath{\elvecsec}{z}				 
\safemath{\nlargefir}{n}
\safemath{\normout}{\gamma}
\safemath{\auxfun}{h}
\safemath{\supp}{\textrm{supp}}

\safemath{\indexa}{\ell}
\safemath{\indexb}{r}
\safemath{\indexc}{i}
\safemath{\indexd}{j}

\safemath{\project}{P}

\def\BibTeX{{\rm B\kern-.05em{\sc i\kern-.025em b}\kern-.08em
    T\kern-.1667em\lower.7ex\hbox{E}\kern-.125emX}}
    
\newcommand{\norm}[1]{\left\lVert #1 \right\rVert}

\newcommand{\PC}[1]{\ensuremath{\!\left(#1\right)}}

\begin{document}

\title{Beam Alignment for the Cell-Free\\mmWave Massive MU-MIMO Uplink}

\author{\IEEEauthorblockN{Jannik Brun, Victoria Palhares, Gian Marti, and Christoph Studer}\\
\textit{Department of Information Technology and Electrical Engineering, ETH Zurich, Zurich, Switzerland} \\ \textit{e-mail: brunj@student.ethz.ch, vmenescal@ethz.ch, gimarti@ethz.ch, and studer@ethz.ch}\\
\thanks{The work of VP, GM, and CS was supported in part by an ETH Research Grant. The work of CS was supported in part by ComSenTer, one of six centers in JUMP, a SRC program sponsored by DARPA. The work of CS was supported in part by the U.S.\ NSF under grants CNS-1717559 and ECCS-1824379.}}

\maketitle

\begin{abstract}
Millimeter-wave (mmWave) cell-free massive multi-user (MU) multiple-input multiple-output (MIMO)
systems combine the large bandwidths available at mmWave frequencies with the improved coverage
of cell-free systems. 
However, to combat the high path loss at mmWave frequencies, user equipments (UEs) must form 
beams in meaningful directions, i.e., to a nearby access point (AP). At the same time, multiple UEs
should avoid transmitting to the same AP to reduce MU interference. 
We propose an interference-aware method for beam alignment (BA) in the cell-free mmWave massive MU-MIMO uplink. In the considered scenario, the APs perform full digital receive beamforming while the UEs perform analog transmit beamforming. 
We evaluate our method using realistic mmWave channels from a commercial ray-tracer, 
showing the superiority of the proposed method over omnidirectional transmission as well as 
over methods that do not take MU interference into account. 
\end{abstract}

\section{Introduction}
Millimeter-wave (mmWave) cell-free massive multi-user (MU) multiple-input multiple-output (MIMO) is expected to be a key to reliable high-throughput wireless connectivity: mmWave frequencies provide large portions of contiguous bandwidth \cite{Rappaport2013} and cell-free systems promise improved coverage even under the notoriously challenging propagation characteristics at mmWave frequencies~\cite{alonzo2017cell}. 
However, these advantages come at the prize of formidable engineering challenges. 
mmWave communication has thus been an active research topic for a number of years \cite{Rappaport2013, heath2016overview, xiao2017millimeter} and recently found its way into commercial products \cite{iphone}. 
While cell-free massive MU-MIMO has also been an area of active research \cite{ngo2017, interdonato2019ubiquitous, buzzi2017cell}, the bulk of studies has been devoted to sub-6-GHz communication and typically assumes single-antenna transceivers both for the user equipments (UEs) as well as for the access points (APs) \cite{ngo2017, buzzi2017cell, interdonato2019ubiquitous}. 

Even in the mmWave cell-free literature, the UEs are often assumed to have a single antenna only \cite{kim2021energy, wang2021millimeter, femenias2019cell, guo2021robust, xue2021, yetis2021}, or, if they have multiple antennas, they do not perform any kind of beamforming with these \cite{alonzo2017cell, alonzo2019energy} (with the exception of \cite{buzzi_beam_2021}). 
However, the high path loss at mmWave frequencies is expected to require UE-side beamforming even in cell-free systems: Rather than aimlessly transmitting the uplink signal in all directions, a UE should form a beam that is aligned to a nearby AP. 
The challenge of determining the optimal beam direction
in which a UE should transmit is exacerbated by the fact that multiple UEs should
avoid transmitting to the same (few-antenna) AP to mitigate multi-user (MU) interference.

\subsection{Contributions}
We consider the problem of beam alignment (BA) in the uplink of cell-free mmWave massive MU-MIMO systems.
In the considered scenario, the UEs are able to perform analog~beamforming 
and the APs of the cell-free system are able to perform digital beamforming. 
In contrast to what is customary in the cell-free literature, we do not assume that the number of APs is much
larger than the number of UEs served \cite{song_joint_2021}, as such an assumption depends on a high pervasiveness of wireless infrastructure. 
We propose a multi-user interference-aware method for BA in the uplink of cell-free mmWave massive MU-MIMO. 
The proposed method is complemented with an efficient block-sparse channel estimation (CHEST) method and a centralized linear minimum mean square error (LMMSE) data detector. 
The resulting scheme is evaluated using realistic mmWave channels from a commercial ray-tracer \cite{remcom}, and with realistic transmit powers and noise figures. 
Our results show that omnidirectional transmission at the UEs leads to inferior performance compared to the proposed scheme. 
In ablation studies we show that the limited capabilities of the UEs compared to fully digital beamforming seem to be of little importance, whereas the interference-awareness of the BA algorithm contributes significantly to performance. 
These findings suggest that the problem of BA in dense cell-free mmWave networks should be considered from a global perspective while the precise radio-frequency (RF) architecture of the UEs is comparably less important. 

\subsection{Relevant Prior Art}
Although there exists a substantial amount of literature on cell-free systems with mmWave frequencies \cite{femenias2019cell, kim2021energy,alonzo2017cell, alonzo2019energy, wang2021millimeter, buzzi_beam_2021, guo2021robust,xue2021,yetis2021}, 
most of these works do either not consider beamforming at all \cite{kim2021energy}, or they consider beamforming only at the APs \cite{femenias2019cell, guo2021robust, xue2021, yetis2021, alonzo2017cell, alonzo2019energy, wang2021millimeter}. References \cite{femenias2019cell, guo2021robust, xue2021, yetis2021} consider APs with different variants of hybrid beamforming and single-antenna UEs, while \cite{wang2021millimeter} considers APs with analog beamforming and single-antenna UEs. References \cite{alonzo2017cell, alonzo2019energy} assume APs with hybrid beamforming and multi-antenna UEs, but the UEs do not use beamforming (the multiple antennas are only used for increased array gain). 
To our knowledge, the only existing work that considers the problem of beamforming or beam alignment on the UE side is \cite{buzzi_beam_2021}, which proposes a scheme in which the UEs simultaneously estimate the direction of arrival (and departure) of the strongest beam. This paper also devises an algorithm that partitions the APs into sets to minimize the interference between APs which use non-orthogonal transmit patterns.

In stark contrast with previous work, we propose an interference-aware UE BA method for the uplink of cell-free mmWave massive MU-MIMO systems, and taking into account multi-antenna APs and UEs. Unlike other works, we consider a scenario in which there are more UEs than APs in the same time-frequency resource, which is of practical importance. Additionally, in opposition to the largely assumed hybrid architectures, we study the situation where UEs rely on analog beamforming and APs are fully digital \cite{bjornson2020}.

\subsection{Notation}
Upper case and lower case bold symbols denote matrices and vectors, respectively. 
The $N\times N$ identity matrix is given by~$\bI_N$.
$\bA^T$ and $\bA^H$ are the transpose and conjugate transpose of $\bA$, respectively; 
and $\bA_{l,k}$ is the matrix in the $l$th row block and $k$th column block of the matrix~$\bA$. 
$\text{diag} \PC{\bma_1,\dots,\bma_N}$ is a block-diagonal matrix whose diagonal blocks are $\bma_1,\dots,\bma_N$.
The Euclidean and Frobenius norms are $\norm{\cdot}$ and $\frobnorm{\cdot}$, respectively. 

\section{System Model}
\begin{figure}[tp]
\centering
\includegraphics[width=\columnwidth]{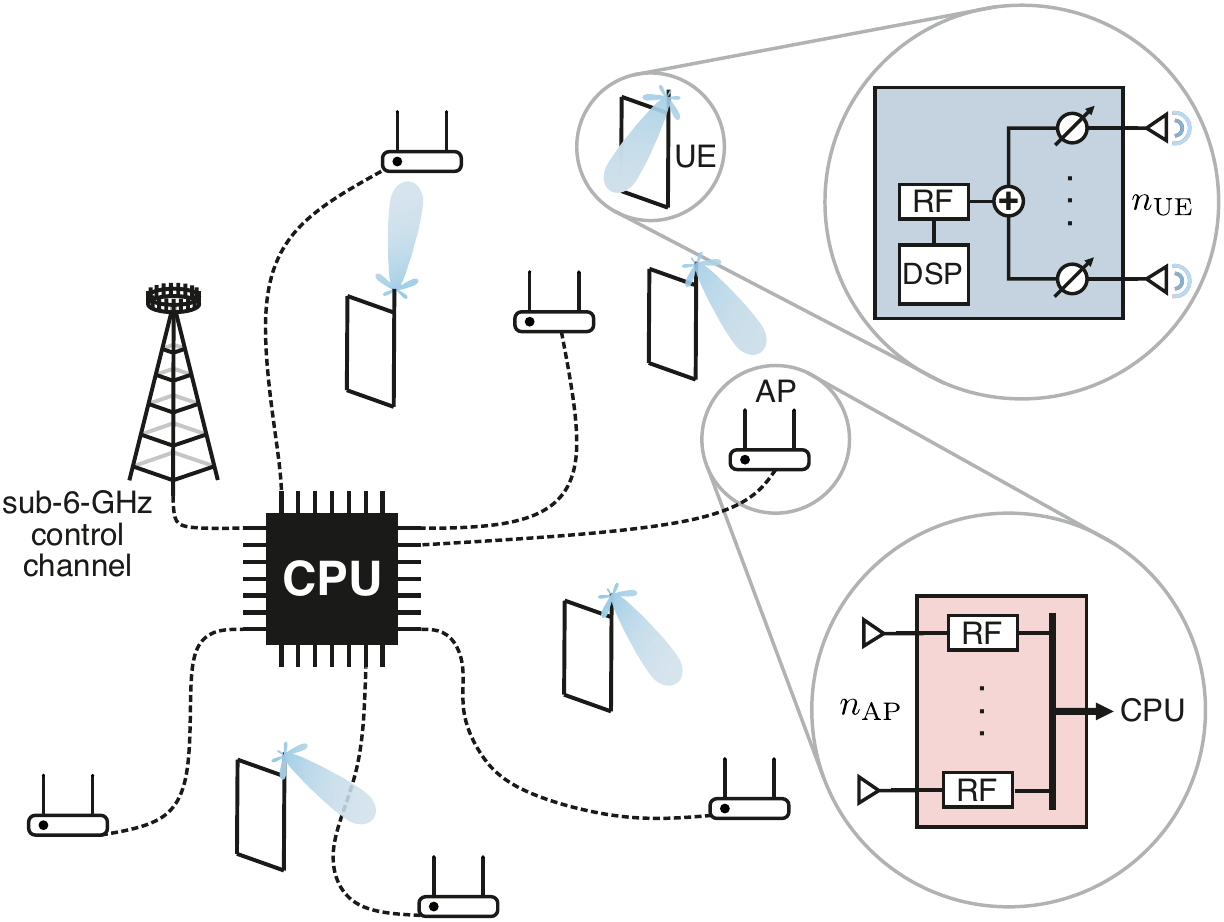}
\vspace{-3mm}
\caption{Visualization of the considered cell-free scenario. The UEs use a single RF chain to load the antennas via phase shifters. The APs are fully digital \cite{bjornson2020}.}
\label{fig:beamformer}
\end{figure}

We consider the uplink of a cell-free mmWave massive MU-MIMO system with $L$ locally distributed access points (APs) that are connected to a central processing unit (CPU) and jointly serve $K$ UEs, see \fref{fig:beamformer}. 
Such a scenario could correspond for instance to a shopping mall or a university campus.
The APs and UEs each have $n_\text{AP}$ and $n_\text{UE}$ antennas respectively, so that the total number of receive and transmit antennas is $N_R=L n_{\text{AP}}$ and $N_T=Kn_{\text{UE}}$, respectively.
The UEs are assumed to have a single radio-frequency (RF) chain and thus rely on analog beamforming; 
the APs are assumed to be fully digital, i.e., every antenna is connected to its own~RF~chain.\footnote{
This is not unrealistic since the APs consist of only few (e.g., four) antennas.}

We assume a frequency-selective mmWave channel which~is divided into $n_{\text{sc}}$ frequency-flat subcarriers using orthogonal frequency-division multiplexing (OFDM). The per-subcarrier input-output relation between all UEs and APs is expressed as
\begin{align}
    \bmy = \bH \bP \bms + \bmn,
\end{align}
where we omit the subcarrier index for ease of notation. 
Here, $\bmy\in\opC^{N_R}$ is the receive vector, $\bH\in\opC^{N_R\times N_T}$ is the channel matrix, 
$\bP\in \opC^{N_T\times K}$ is the UE beamforming matrix, $\bms\in\opC^K$ is the vector of UE transmit symbols with per-symbol energy $E_s$, 
and $\bmn\in\opC^{N_R}$ is i.i.d. circularly-symmetric complex Gaussian noise with variance~$N_0$ per complex entry. 
The channel matrix has the following block structure, 
\begin{align}
\bH = \begin{bmatrix}
\bH_{1,1} & \cdots & \bH_{1,K} \\
\vdots & & \vdots \\
\bH_{L,1} & \cdots & \bH_{L,K}
\end{bmatrix}, \label{eq:H_matrix}
\end{align}
where $\bH_{\ell,k}\in \opC^{n_{\text{AP}} \times n_{\text{UE}}}$ is the channel matrix from the $\ell$th AP to the $k$th UE. We will also use $\bH_k=[\bH_{1,k}^T, \dots, \bH_{L,k}^T]^T \in \opC^{N_R \times n_{\text{UE}}}$ to denote the $k$th column block of $\bH$. 
Any effects of power control are absorbed into the channel matrix. 
The UE beamforming matrix $\bP$ is block-diagonal, $\bP=\text{diag}(\bmp_1,\dots,\bmp_K)$, with diagonal blocks $\bmp_k \in \opC^{n_{\text{UE}}}, k=1,\dots,K$.
The block-diagonal structure reflects the fact that the UEs cannot transmit cooperatively. 
We assume UEs with a single RF chain and analog phase shifters, so that the codebook of possible beams is
\begin{align}
    \setP = \Big\{& \textbf{p}(\phi_b)= \frac{1}{\sqrt{n_{\text{UE} }}}[1, e^{-j\pi \cos \phi_b}, \dots, \, e^{-j\pi \cos\phi_b (n_{\text{UE}}-1)} ]^T, \nonumber\\
    &\phi_b\in\Big\{0,\frac{\pi}{B},\dots,\frac{B-1}{B}\pi\Big\}
    \Big\}, \label{eq:setP}
\end{align}
where $|\setP|=B$ is the codebook size. 
Note that, since we use analog beamforming, the beamforming matrix $\bP$ is identical for all subcarriers.

As in \cite{buzzi_beam_2021}, we assume the existence of a sub-6-GHz control channel which can be used for control signaling from the CPU to the UEs. 
However, compared to \cite{buzzi_beam_2021}, our scheme requires significantly less control signaling overhead: It only requires the CPU to communicate to each UE the codebook index of its selected beam, and requires no control signals from the UEs. 

\subsection{Communication Scheme} \label{sec:scheme}
Multi-user interference-aware BA requires knowledge of the channel matrix in \eqref{eq:H_matrix}
or some surrogate thereof. Knowledge of the channel matrix in wireless systems is typically acquired
through the transmission of pilots, which leads to a chicken-or-egg problem: 
On one hand, if the UEs transmit pilot symbols without using appropriate beamforming vectors, the APs will 
only pick up a very weak signal which does not allow reliable CHEST.\footnote{Because of channel reciprocity, the same would also hold if the pilots were transmitted by the APs instead of the UEs.} 
On the other hand, some knowledge of the channel matrix is necessary for the UEs to select appropriate~beams.

We solve this conundrum using a two-stage scheme: 
In the first stage, the UEs transmit non-orthogonal pilots in different directions to ensure sufficient receive signal strength
for at least some of the directions. Based on these pilots, the CPU then performs a coarse estimate 
of the channel matrix which it then uses for BA. Note that this estimate of the channel matrix
only needs to be accurate enough to determine the angles of strong receptivity.
In the second stage, the UEs fix their beamformers to the selected beams, which are communicated to them via the available sub-6-GHz control channel. 
The UEs transmit orthogonal pilot sequences used by the CPU to perform a high-quality estimate of the 
compound of the channel matrix and the beamforming matrix $\bH\bP$. This estimate is then used for data detection, 
which in this paper we do using centralized LMMSE equalization $\hat{\bms} = \bW^H \bmy$
with
\begin{align}
    \bW = \bH\bP \left((\bH\bP)\sp{H}\bH\bP+ \frac{N_0}{E_s} \bI_K \right)^{-1}. \label{eq:lmmse}
\end{align}
Many other detection methods, including decentralized ones, would of course be possible and are left for future work.

We will now first describe the BA procedure before explaining the pre-beam-alignment (pre-BA) and 
post-beam-alignment (post-BA) CHEST steps that precede and follow it, respectively. 

\section{Beam Alignment} \label{sec:beam_selection}
As stated in \fref{sec:scheme}, BA requires knowledge of the channel matrix (or some suitable surrogate thereof). It is important to remember that here we consider frequency-selective mmWave channels that can be divided into frequency-flat subcarriers. Ideally, BA would be applied to each of the subcarriers independently. However, since the UEs use analog beamforming, they can only form a single beam, which is then used for all subcarriers. 
We therefore perform BA based on the strongest subcarrier channel matrices per UE and AP, i.e., based on a surrogate matrix $\bar{\bH}\in\opC^{N_R\times N_T}$ consisting of the
blocks $\bar{\bH}_{\ell,k} = \argmax_{v=1,\dots,n_\text{sc}} \big\|\bH^v_{\ell,k}\big\|_\textnormal{F}^2$, 
where $v$ is the subcarrier~index.

Even when using such a frequency-flat surrogate, optimal BA is typically not practical for common optimization objectives such as achievable sum-rate or max-min achievable rate since it requires the optimization over a combinatorial set. 
For instance, for max-min achievable rate optimal BA, we would be required to solve the following optimization problem: 
\begin{align}
    \max_{(\bmp_1, \dots, \bmp_K) \in \setP^K}~\min_{k=1,\dots,K}~ \textit{SINR}_k (\bmp_1, \dots, \bmp_K),
\end{align}
where $\textit{SINR}_k (\bmp_1, \dots, \bmp_K)$ is the post-equalization 
signal-to-interference-plus-noise ratio (SINR) of the $k$th UE given the beamforming vectors $\bmp_1, \dots, \bmp_K$:
\begin{align}
  \textit{SINR}_k (\bmp_1, \dots, \bmp_K) = 
  \frac{|\bmw_k^H \bH_k \bmp_k |^2}{\sum_{\substack{k'\neq k}} |\bmw_k^H \bH_{k'} \bmp_{k'}|^2 + \No\norm{\bmw_k}_2^2},
  \label{eq:sinr}
\end{align}
with $\bmw_k$ being the $k$th column of \eqref{eq:lmmse}. Solving this optimization problem through an exhaustive 
search for $K=32$ UEs and a codebook size of $B=16$ would amount to $B^K\approx10^{38}$ evaluations of the 
objective in \eqref{eq:sinr}. Clearly, this is not feasible in practice. 
We therefore now outline some more practical methods for BA. 

\subsection{Interference-Unaware Digital Beam Alignment} \label{sec:digital_beam_selection}
We start by discussing the well-known eigenbeamforming method \cite{tse2005fundamentals}, which in our case is not applicable because it would require digital beamforming, but which we will later use as an evaluation baseline. 
If multi-user interference were negligible (and the UEs could perform full digital beamforming), 
one could simply optimize the (pre-equalization) signal-to-noise ratio (SNR) for every user, by solving
\begin{align}
    \max_{\bmp} \frac{E_s\|\bH_k \bmp\|^2}{N_R\No}~\text{s.t. }\|\bmp\|=1, ~k=1,\dots,K, \label{eq:max_snr}
\end{align}
which yields the leading (=\,belonging to the largest singular value) right-singular vector of the per-UE channel matrix $\bH_k$.

\subsection{Interference-Unaware Analog Beam Alignment} \label{sec:analog_IU}
In the considered scenario, the UEs can only form beams from the finite set \eqref{eq:setP} and so 
in general will be unable to beamform the leading right-singular vector of $\bH_k$. However, if multi-user interference 
is negligible or ignored, we can still solve \eqref{eq:max_snr} to maximize the (pre-equalization) SNR for every
UE, except that we now optimize over the codebook of possible beams $\setP$ instead of the set $\{\bmp:\|\bmp\|=1\}$.
Since this optimization problem can be solved individually per UE, we can solve it using an exhaustive search 
over $KB$ possibilities. 

\subsection{Interference-Aware Analog Beam Alignment} \label{sec:analog_IA}
As we will see in the \fref{sec:results}, ignoring MU interference results in suboptimal performance. 
For this reason, our proposed BA algorithm takes MU interference into account. 
We have already seen that maximizing joint optimization objectives, such as max-min-optimal achievable rate is
impractical. For this reason, we use an algorithm in which the UEs greedily optimize
their individual post-equalization SINR, given the currently selected beams of the other UEs. 
Such an algorithm can be succinctly motivated by a German idiom:\footnote{``Wenn jeder an sich denkt, ist an alle gedacht.''} ``When everyone thinks of themselves, everyone is taken care of.''
The method operates as follows:
First, we perform a singular value decomposition (SVD) for each per-UE channel matrix $\bH_k$, $k=1,\dots,K$. 
We then sort the UEs in descending order according to the largest singular value $\sigma_k$ of their per-UE channel matrix $\bH_k$.
Assume now that the UEs are ordered in this way, i.e., $\sigma_k>\sigma_i$ for $i>k$. 
Starting from an empty set of UEs, we then iteratively select the beam for the $k$th UE which maximizes the post-equalization 
SINR, where the interference-term only takes into account the interference of those UEs whose beams have already been selected: 
\begin{align}
    \bmp_k = \argmax_{\tilde{\bmp}_k} \frac{|\bmv_k^H \bH_k \tilde{\bmp}_k |^2}{\sum_{k'<k} |\bmv_k^H \bH_{k'} \bmp_{k'}|^2 + \No\norm{\bmv_k}_2^2}, \label{eq:ascending}
\end{align}
for $k=1,\dots,K$.  
In \fref{eq:ascending}, $\bmv_k$ is the $k$th column of the LMMSE equalization matrix as in \eqref{eq:lmmse}, but restricted to only 
the first $k$ UEs, i.e., to those UEs whose beam has already been selected or is currently being selected. 
This means that $\bmv_k$ is a function of $\bmp_1,\dots, \bmp_{k-1}$ and $\tilde{\bmp}_k$, 
as well as $\bH_1,\dots,\bH_k$.

Then, in a second round, we use a coordinate descent approach where every UE has the chance to re-adjust its beams to adapt to the beams of the UEs that were selected later
by taking into account the interference of \emph{all} UEs according to their curently selected beams.
This round iterates in reverse order, $k=K,\dots,1$,
\begin{align}
    \bmp_k = \argmax_{\tilde{\bmp}_k} \frac{|\bmv_k^H \bH_k \tilde{\bmp}_k |^2}{\sum_{k'\neq k} |\bmv_k^H \bH_{k'} \bmp_{k'}|^2 + \No\norm{\bmv_k}_2^2}, \label{eq:descending}
\end{align}
where $\bmv_k$ now denotes the $k$th column of the LMMSE equalization matrix \eqref{eq:lmmse} considering \emph{all} UEs, 
i.e., $\bmv_k$ is now a function of $\bmp_1,\dots,\tilde{\bmp}_k,\dots,\bmp_K$, as well as $\bH_1,\dots,\bH_K$.

\section{Channel Estimation} \label{sec:chest}
The previous section on BA assumed that the channel matrix $\bH$ is known for every subcarrier. 
We now describe our proposed method to estimate the channel with sufficient accuracy 
for the BA procedure. 

\subsection{Pre-Beam-Alignment Channel Estimation (Pre-BA CHEST)} \label{sec:pbs_chest}
To perform BA based on the strongest subcarrier, we would in principle have to estimate the 
channel matrix for every subcarrier individually. However, to keep complexity at bay, and 
since large-scale fading properties do not change dramatically between adjacent subcarriers, 
we only estimate the channel matrix for a subset of subcarriers which are spaced uniformly
over the total number of used subcarriers. 

We now describe the CHEST in per-subcarrier fashion, where we again omit the subcarrier index 
for brevity.
We estimate the channel matrix at the APs using pilot signals transmitted at the UEs for $n_\textit{pilot}$ time slots, 
i.e., we observe
\begin{align}  
    \bY = \bH\bB + \bN,
\end{align}
where the matrix $\bB \in \opC^{N_T\times n_\textit{pilot}}$ is the compound of the UE transmit symbols and the beams they form while transmitting them. 
Assuming that the matrix $\bB$ (which we call the beam-pilot matrix) is known, the APs estimate
$\bH\in \opC^{N_T\times N_R}$ based on $\bY \in \opC^{N_R\times n_\textit{pilot}}$. 
Note that, since this happens before the BA phase, the beams which can be used by the UEs in $\bB$
need to be predetermined. 
To explore the channels' characteristics as fully as possible, and to make sure that we get adequate signal 
receive power at least during some of the time slots of $\bY$, the UEs essentially sweep through the beam codebook. 
The $k$th UE transmits the per-UE beam-pilot matrix $\bB_{(k,\cdot)}\in \opC^{n_\text{UE}\times n_\textit{pilot}}$, which is the $k$th row block of~$\bB$, 
and whose columns contain all beams $\bmp\in\setP$.\footnote{This means that the UE pilot symbols for 
the pre-BA CHEST all equal $1$.}
If all of the UEs would transmit their beam pilots simultaneously, we would have the most efficient CHEST
scheme with $n_\textit{pilot}=B$. However, depending on the number of UEs $K$ and the size of the beam codebook $B$, 
the matrix $\bB$ can become very tall, which results in degenerate estimates of the channel matrix $\bH$. 
Conversely, if only one UE at a time transmits its pilot beams, CHEST is much easier but
the scheme becomes time consuming, with $n_\textit{pilot}=BK$~time~slots.

To develop a scheme which is both sufficiently accurate and efficient, we use a clustering 
scheme, where the $K$~UEs are randomly clustered into $C$ equisized clusters: All UEs belonging to 
the same cluster transmit their pilot beams simultaneously, while the different clusters transmit their 
pilot beams sequentially, resulting in $n_\textit{pilot}=BC$.

We also use an informative prior that exploits the block sparsity of the channel matrix $\bH$
which results from the distributed nature of the APs and the UEs in combination with the high directivity of 
mmWave channels \cite{Rappaport2013}.
Specifically, we estimate $\bH$ by solving the convex optimization problem
\begin{align}
    \est{\bH}= \argmin_{\bH \in \opC^{N_R\times N_T}} \frac{1}{2}\norm{\bY-\bH\bB}_\text{F}^2 + \mu \norm{\bH}_\text{BS}, 
    \label{eq:est_H}
\end{align}
where $\mu$ is a regularization parameter and we use the block sparsity promoting prior \cite{cotter2005sparse}
\begin{align}
    \norm{\bH}_\text{BS} = \sum\sb{\ell=1}^L \sum\sb{k=1}^{K}  \norm{\bH_{\ell,k}}\sb{\text{F}}.
\end{align}
The optimization problem \eqref{eq:est_H} can be efficiently solved using forward-backward-splitting (FBS) 
\cite{goldstein2014field}. 

\subsection{Post-Beam-Alignment Channel Estimation (Post-BA CHEST)}
Using the estimates $\est{\bH}$ in \eqref{eq:est_H} of the channel matrices for the different subcarriers, the CPU then performs BA 
as detailed in \fref{sec:beam_selection} and communicates the selected beams to the UEs using the 
sub-6-GHz control channels. The UEs then fix their beams accordingly in the form of the 
beamforming matrix $\bP$. With these beams now being fixed, we effectively have a compound channel matrix
$\bH\bP \in \opC^{N_{R}\times K}$ for every subcarrier. 
Since the beams $\bP$ have been appropriately selected, we can estimate this compound channel (for \textit{every} subcarrier)
with high accuracy for reliable data detection. We do this by letting the UEs transmit pilot sequences
which together form a Hadamard matrix. The APs then estimate $\bH\bP$ using least squares
estimation. 

\section{Numerical Results} \label{sec:results}
We now evaluate the performance of our proposed scheme in comparison with a number of baselines.
\subsection{Simulation Setup}
We simulate a local cell-free communication scenario in which $K=32$ UEs, 
each with $n_\text{UE}=8$ antennas, communicate with $L=16$ APs, each with $n_\text{AP}=4$ antennas.
We use a carrier frequency of 28 GHz and a bandwidth of 1 GHz divided into $n_{\text{sc}} = 2048$ orthogonal subcarriers. For the pre-BA CHEST procedure, we estimate the channels of the subcarrier indices $v \in \{1,\,229,\, 456,\, 684,\, 911,\, 1139,\, 1366,\, 1594,\, 1821,\, 2048 \}$ and considering $C = 8$ equisized clusters. 
The BA step is performed based on the strongest subcarrier channel matrices per UE and AP,
as detailed in \fref{sec:beam_selection}.
The UEs use a QPSK transmit constellation and transmit at a maximum power of 20\,dBm
using $\pm\,3$dB power control.
To model realistic RF hardware, the APs have a noise figure of 7\,dB \cite{rundao2020}.

\begin{figure}[tp]
\centering
\includegraphics[width=0.75\columnwidth]{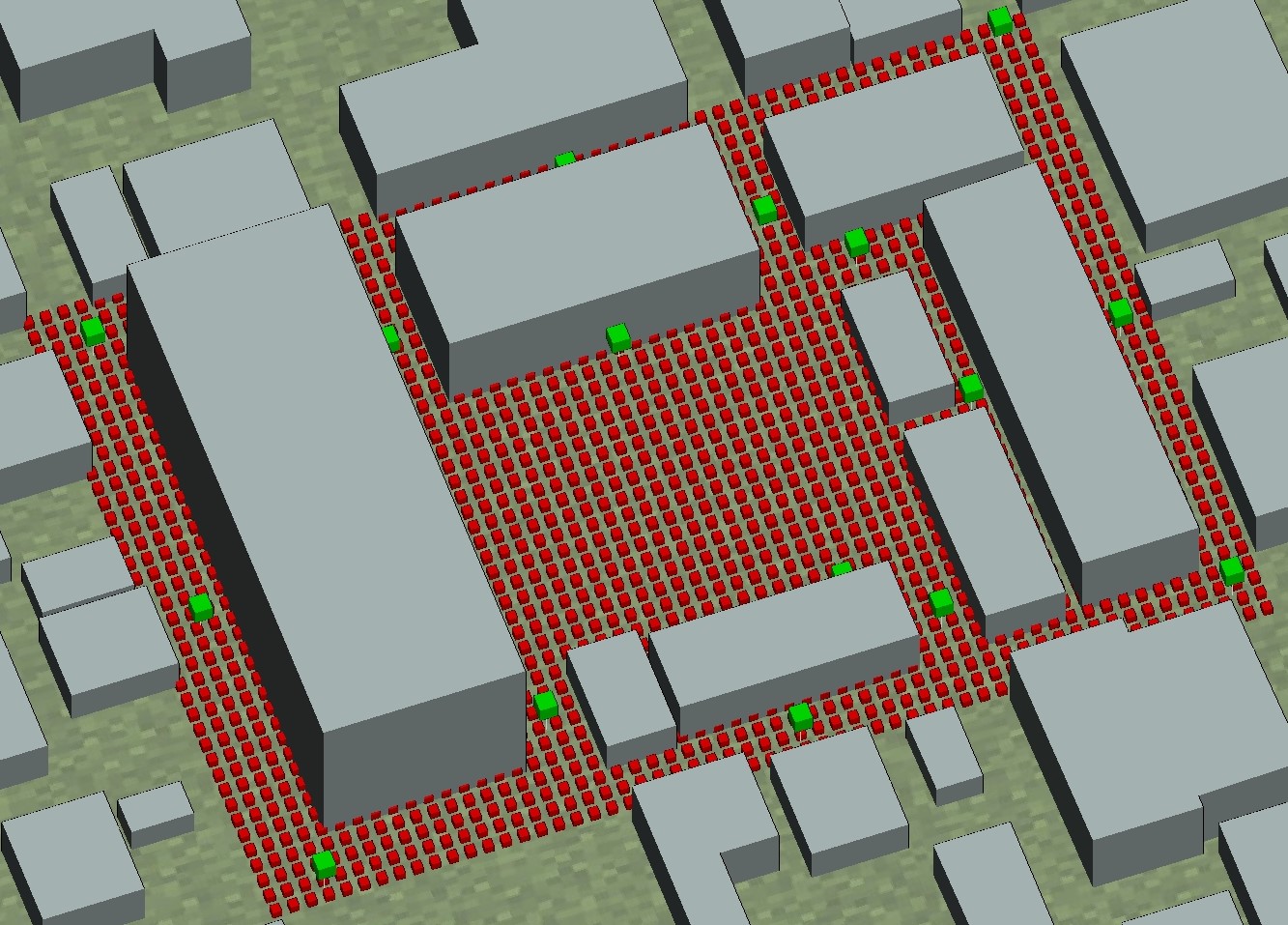}
\caption{Simulated scenario with $1336$ possible UE locations (red), each with $n_\text{UE}=8$ antennas, and $L=16$ APs (green), each with $n_\text{AP}=4$ antennas.}
\label{fig:fig_sim}
\end{figure}

The wireless channels are simulated using Wireless InSite from Remcom \cite{remcom}. The setup, which is depicted in 
\fref{fig:fig_sim}, consists of a $200\,\text{m}\times280\,\text{m}$ area and contains 1336 possible UE locations
(visualized in \fref{fig:fig_sim} as red squares) which are arranged on a grid. To simulate random user orientations, we also consider four possible angles for each user: $\{0^{\circ},45^{\circ},90^{\circ},135^{\circ}\}$. The position and orientation of the APs are fixed. The APs and UEs are at heights of $12\,$m and $1.65\,$m, respectively. Both APs and UEs have uniform linear arrays with omnidirectional antennas spaced by half a wavelength. 

\subsection{Performance Metrics and Baseline Methods}
As performance metrics we consider the cumulative density function (CDF) or complementary CDF (CCDF) of
the root-mean-squared-symbol error (RMSSE) per UE, the SINR per UE per subcarrier, and the spectral efficiency (SE) per UE (summed over all subcarriers), all considering an LMMSE equalizer. The RMSSE for the $k$th UE is defined as 
\begin{align}   \label{eq:rmsse}
\textit{RMSSE}_k = \sqrt{\frac{\sum_{v=1}^{n_{\text{sc}}} \sum_{t=1}^{T_D}\big|\hat{s}^{(tv)}_{k} - s^{(tv)}_{k}\big|^2}{
\sum_{v=1}^{n_{\text{sc}}} \sum_{t=1}^{T_D}\big|s^{(tv)}_{k}\big|^2 }},
\end{align}
where $T_D$ is the number of time slots and $\hat{s}^{(tv)}_{k}$ and $s^{(tv)}_{k}$ are the estimated and transmitted symbols of the $k$th UE in the $v$th subcarrier in the $t$th time slot.

In our numerical results, we refer to the interference-unaware analog BA from \fref{sec:analog_IU} as ``analog IU'' and to the interference-aware analog BA from \fref{sec:analog_IA} as ``analog IA.'' As the first baseline method, we use ``single antenna,'' where each UE has $n_\text{UE}=1$ antenna, i.e., no BA capabilities. For a fair comparison, the total transmit power of ``single antenna'' UEs is the same as in the other techniques. The second baseline is the interference-unaware digital BA (\fref{sec:digital_beam_selection}), which we refer to as ``digital IU.''

\subsection{Performance Results}
\begin{figure*}[tp]
\centering
\subfigure[]{\includegraphics[height=4.3cm]{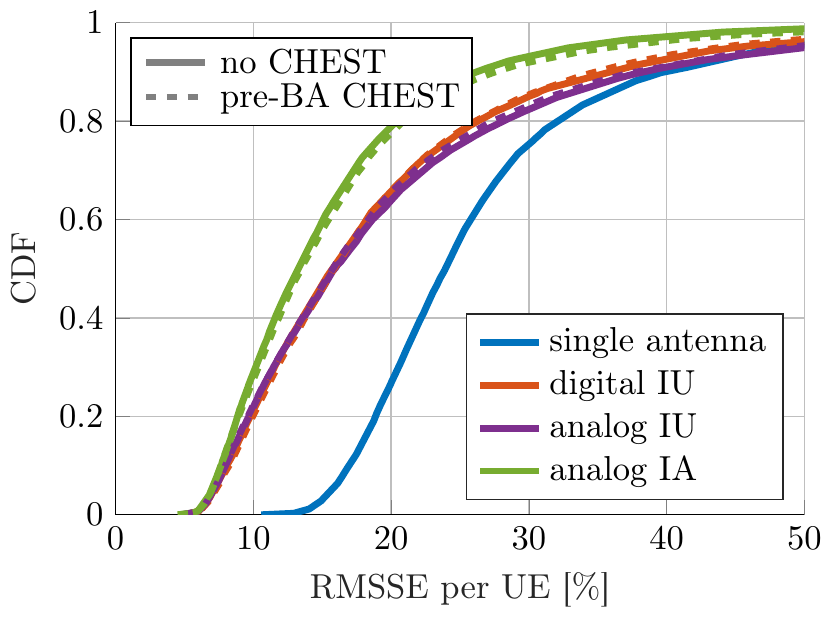}\label{fig:cdf_rmsse}}
\subfigure[]{\includegraphics[height=4.3cm]{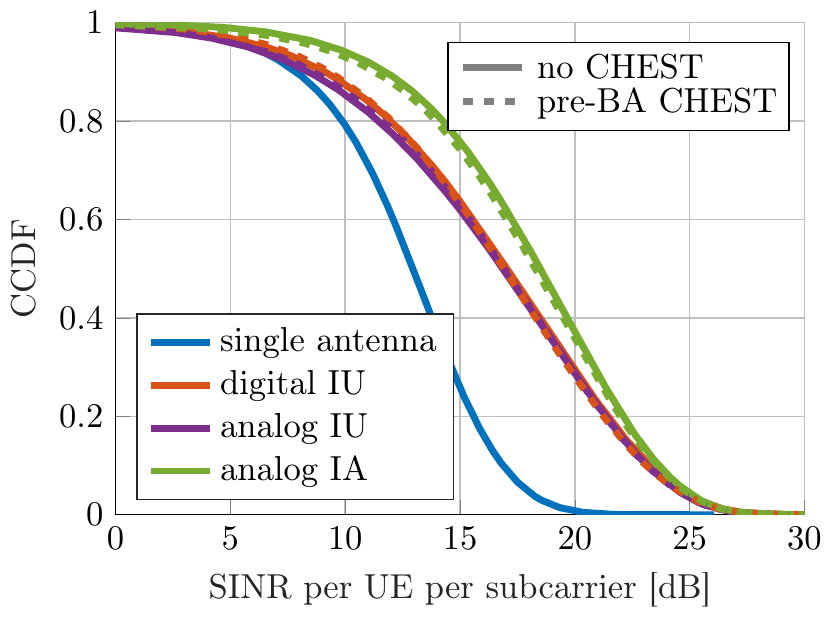}\label{fig:cdf_sinr}}
\subfigure[]{\includegraphics[height=4.3cm]{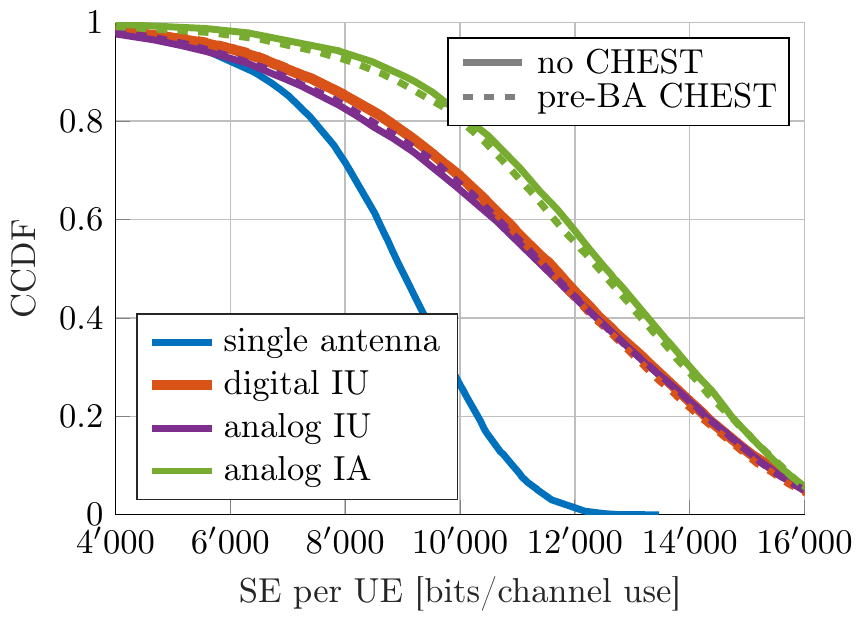}\label{fig:cdf_rate}}
\vspace{-1mm}
\caption{Subfigures (a), (b), and (c) show the CDF or CCDF of the RMSSE per UE, the SINR per UE per subcarrier, and the SE per UE, respectively, for ground truth channel knowledge (solid) or pre-BA CHEST in combination with perfect post-BA CHEST (dashed).}
\vspace{-3mm}
\label{fig:cdf}
\end{figure*}

\begin{figure*}[tp]
\centering
\subfigure[]{
\includegraphics[height=4.3cm]{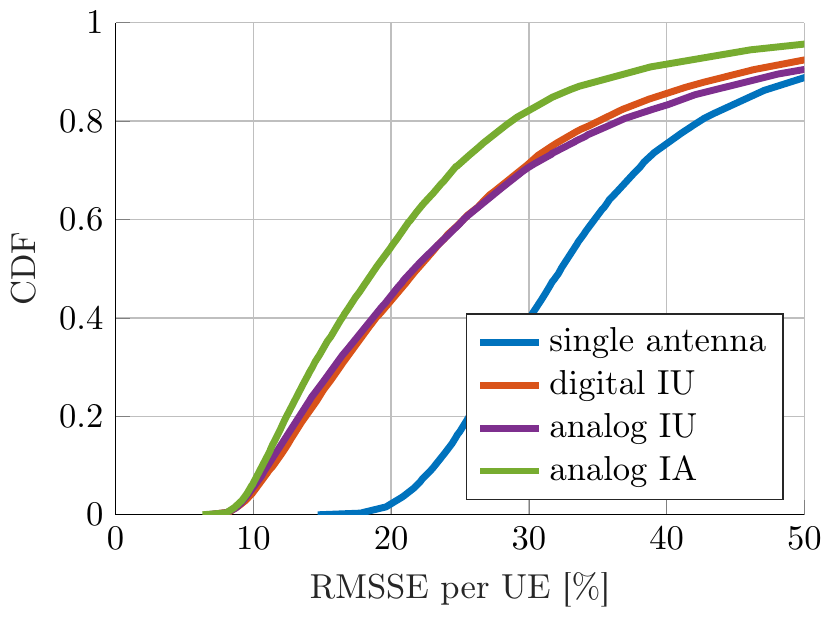}\label{fig:chest_cdf_rmsse}}
\subfigure[]{
\includegraphics[height=4.3cm]{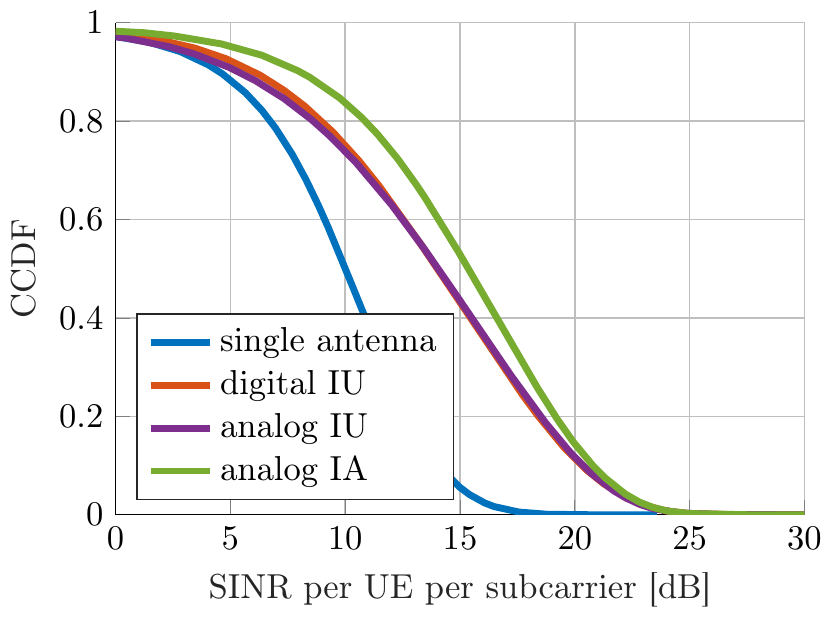}\label{fig:chest_cdf_sinr}}
\subfigure[]{
\includegraphics[height=4.3cm]{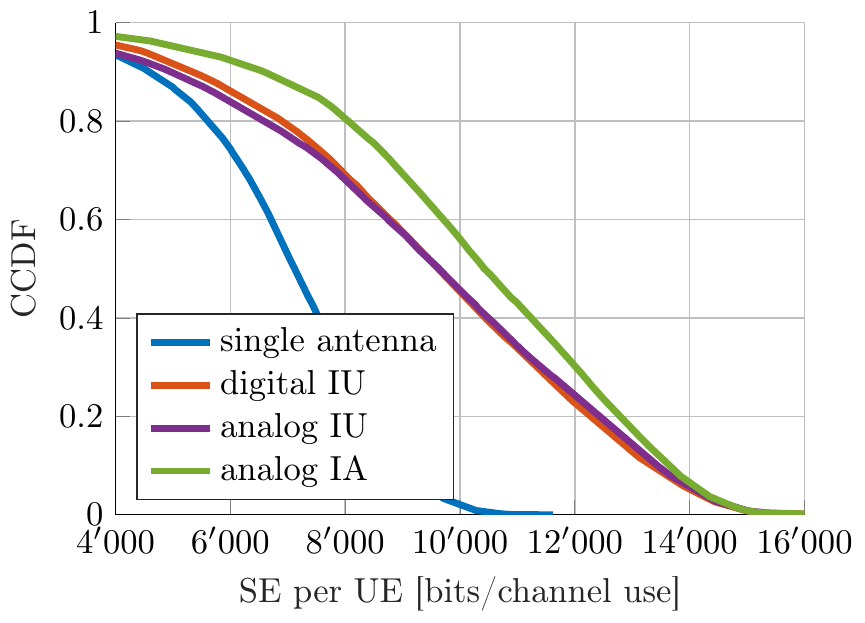}\label{fig:chest_cdf_rate}}

\vspace{-1mm}
\caption{Subfigures (a), (b), and (c) show the CDF or CCDF of the RMSSE per UE, the SINR per UE per subcarrier, and the SE per UE, respectively, in the absence of ground truth channel knowledge, i.e., using both pre-BA and post-BA CHEST.}
\vspace{-3mm}
\label{fig:chest_cdf}
\end{figure*}

\fref{fig:cdf} shows BA results which consider ground-truth channel knowledge; \fref{fig:chest_cdf} shows results where no such knowledge is available and hence we estimate the channel as described in \fref{sec:chest}. To prove the efficacy of our proposed pre-BA CHEST scheme from \fref{sec:pbs_chest}, 
\fref{fig:cdf} also contains results where we combine pre-BA CHEST for BA with ground truth channel knowledge for data equalization (dashed curves). 
The fact that the results where we perform \mbox{pre-BA} CHEST are virtually identical to the results where perfect channel knowledge is available also during the BA step proves that our pre-BA CHEST procedure solves its task nearly optimally. Any performance deterioration from \fref{fig:cdf} to \fref{fig:chest_cdf} is due to the loss incurred by the post-BA CHEST for data detection, where we only used a plain vanilla least squares channel estimator.

The performance hierarchy between the methods is very similar for all performance metrics, 
and regardless of whether channel knowledge is available or has to be estimated. 
The ``single-antenna'' method has the worst performance among all considered approaches. 
This shows that omnidirectional single-antenna UEs suffer from significant performance losses in cell-free mmWave systems compared to beamforming-capable UEs with adequately selected beams. 
The ``digital IU'' method and its analog counterpart ``analog IU''
perform substantially better. 
Notably, ``analog IU'' is almost as good as ``digital IU,'' suggesting that the more limited expressiveness of the beams that can be formed by the ``analog IU'' compared to ``digital IU'' are not negatively affecting the performance. 
Finally, the ``analog IA'' achieves by far the best performance. 
The difference between ``analog IA'' and the interference-unaware methods is larger for the weakest UEs. The spectral efficiency of the 10\% weakest UEs is essentially $1.3\times$ better for the interference-aware method as for the interference-unaware methods. This shows that optimal BA of dense cell-free mmWave networks requires algorithms that select the beams for the UEs jointly and take multi-user interference into account. 
Contrastingly, the precise RF architecture and resulting beamforming expressivity of the UEs seem to be comparably less important. 

\section{Conclusions}
We have proposed an interference-aware beam alignment method together with a channel estimation scheme for cell-free mmWave massive MU-MIMO systems. 
Our results have shown that multi-user interference-aware beam alignment consistently outperforms methods
that do not take interference into account, which in turn outclass omnidirectional transmission without beamforming. 
We have also shown the efficacy of our proposed pre-beam-alignment channel estimation method in comparison to a hypothetical scenario where ground truth channel knowledge is available. 
For future work we highlight the combination of our methods with decentralized data detection schemes as well as considering the UE beam alignment problem in the mmWave cell-free downlink. 

\bibliographystyle{IEEEtran}
\bibliography{bib/VIPabbrv,bib/confs-jrnls,bib/publishers,bib/VIP}

\vspace{12pt}

\end{document}